# New Methods for Modal Decomposition in Multi-Mode fibres

S. Blin[1], T. N. Nguyen[1], D. M. Nguyen[1], P. Rochard[1], L. Provino[2], A. Monteville[2],
T. Robin[3], A. Mugnier[4], B. Cadier[3], D. Pureur[4], M. Thual[1], T. Chartier[1]

[1] Laboratoire Foton, CNRS / Université de Rennes 1 (UMR 6082), Enssat, BP 80518, 22305 Lannion cedex, France
[2] Perfos, 11 rue Louis de Broglie, 22300 Lannion, France
[3] iXFiber, Rue Paul Sabatier, 22300 Lannion, France
[4] Quantel, Etablissement R&D Lannion, 4 rue Louis de Broglie, Bât. D, 22300 Lannion, France

**ABSTRACT**

We propose and demonstrate two methods for modal decomposition in multi-mode fibres. Linearly polarized modes propagating in a slightly multi-mode fibre are easily retrieved from intensity measurements at the fibre output surface. The first method is an improvement of the so-called spectrally and spatially imaging technique, which is limited to large-mode-area optical fibers. The second method is a new, simpler and faster solution for the characterization of any kind of optical fibre, thus attractive in comparison to previously reported methods, which are cumbersome, time-consuming and/or limited to large-more-area fibres. Different kinds of multi-mode optical fibres are characterized. A large-mode-area photonic-bandgap fibre, a photonic-crystal small-core non-linear fibre, and a standard index-stepped multi-mode fibre are characterized successfully.

## 1. INTRODUCTION

### 1.1 Context

Multi-mode propagation within an optical fibre is attractive, necessary or problematic depending on the application. For sensing applications, multi-mode fibres are attractive, for example, for their wide evanescent field, which is used to effectively detect chemicals outside the fibre. For high-power applications, multi-mode fibres are necessary to reduce non-linear effects due to their large mode-areas. However, multi-mode propagation deteriorates the beam quality at the output of the fibre, which can be detrimental in some applications such as high power fibre lasers for marking application. For long-distance telecommunications, multi-mode propagation is problematic due to inter-modal dispersion, which strongly limits the maximum data rate. In the last decade, the advent of photonic-crystal fibres[1] offered all-new possibilities for controlling light, such as endlessly single-mode propagation. However, some photonic-crystal fibres often exhibit multi-mode propagation, either due to fibre design or imperfections in the fabrication processes. Therefore, modal analysis of multi-mode fibres is necessary in many applications, in order to identify the existing modes and to determine the power distribution within this set of modes.

### 1.2 State of the art

The most common technique to observe the transverse modes supported by an optical waveguide is to inject light at one extremity of the guide, and to vary the angle of injection while observing the near-field image at the other extremity. Transverse modes discrimination is so possible because these modes have different numerical apertures. However, this technique does not guarantee the observation of all modes individually. Some modes are difficult to excite and a superposition of modes is usually observed at the fibre output. Moreover, this technique does not provide any information about the power distribution within the eigenmode basis for a given injection. Other methods[2] are used to determine this power distribution, but the transverse profiles of eigenmodes have to be known *a priori*, and this prerequisite is not necessarily available for complex and inhomogeneous structures. Recently, Shapira et al.[3] have published a method to determine experimentally the mode profiles using an algorithm based on the observation of near-field and far-field images at the waveguide output. This technique also provides the power distribution within the eigenmode basis, thus offering an interesting solution. The disadvantages though are the complexity of the algorithm and the need to acquire two images (near- and far-fields). Another exciting technique has been demonstrated by Nicholson et al.[4,5] for large-mode-area fibres, namely the $S^2$ imaging, for *spatially and spectrally resolved imaging*. It consists in injecting a broadband source in the fibre and measuring the optical spectrum in every position (x,y) of the output surface of the fibre by collecting light using a single-mode fibre. Spatial accuracy requires the use of a two-objectives U-bench magnification system between the fiber under test and the single-mode fiber. Optical spectra are then processed to retrieve the transverse intensity profile of the linearly polarized (LP) modes[6] along with their optical powers.

In this paper, we propose and demonstrate two alternative solutions to the $S^2$ imaging. The first technique has been used to characterize a large-mode-area photonic-bandgap fibre designed for high-power fibre lasers emitting at 1064 nm. This technique is similar to the one proposed by Nicholson et al., but uses a microlensed fibre for imaging instead of a standard single-mode fibre (along with a magnification system). We will describe a second technique based on the use of a tuneable laser source instead of a broadband source. To the best of our knowledge, this is an original technique offering many advantages such as simplicity and rapidity, due to the fact that the complex two dimensional (2D) scanning of the $S^2$ imaging is no more necessary. The use of a tuneable laser source for modal decomposition has been proposed elsewhere[5] but was combined to the 2D scanning, thus not offering as much interest as the technique proposed here.

## 2. PRINCIPLE

### 2.1 Review of the $S^2$ imaging

The principle of the experiment is presented in Fig. 1 and has been detailed in [4,5]. Light from an optical source is injected in the fibre under test. If this fibre supports propagation of two transverse modes, each eigenmode propagates at its own group velocity, and the two modes experience a phase-shift while propagating along the fibre. If $L$ is the fibre length, $\lambda$ the wavelength, and $\Delta n_g$ the group velocity difference, the phase-shift $\Delta\varphi$ between the two modes at the fibre output is $\Delta\varphi = 2\pi \Delta n_g L / \lambda$. Therefore, interferences occur in every position (x,y) of the fibre output. By scanning the wavelength, one would observe fringes (cf. inset of Fig. 1) whose period is $\Delta\lambda = \lambda^2 / (\Delta n_g L)$. The intensity of both modes at this position (x,y) can be retrieved from these fringes by using a simple algorithm reported in [4,5]. If other eigenmodes propagate in the fibre, interferences will also occur but with a different period as long as $\Delta n_g$ differs, thus allowing to retrieve all the existing modes using Fourier analysis. Finally, it is important to indicate that retrieving mode profiles in this manner is valid as long as most of the optical power is carried mainly by one mode, and that the wavelength span is small enough to suppose that the intensity mode profile does not change significantly with wavelength.

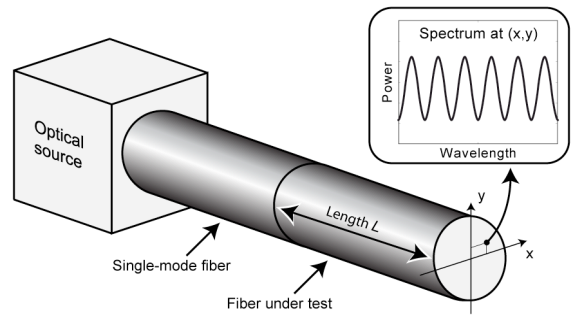

Fig. 1. Modal decomposition principle.

### 2.2 Microlensed fibre imaging for the $S^2$ technique

Nicholson et al.[4,5] proposed to use a broadband source and to scan the fibre output plane by interrogating the magnified image of the fibre end with a single-mode fibre. We used a broadband source as well for this first technique, but we propose to avoid the magnification setup by simply using a microlensed optical fibre. As shown in Fig. 2(a), the microlensed fiber has a working distance of 100 µm and a spot-diameter of 3 µm (at 1550 nm). This homemade fibre is used as a probe to collect light exiting the fibre under test at every position of its output face. Closed-loop piezoelectric transducers are used for scanning with sub-micrometer accuracy. Optical spectra are acquired at every position, and then processed to retrieve the mode profiles of the fibre under test. Scanning a 60x60-µm square area with a 2-µm step has been done in 48 hours due to the slow acquisition time of the optical spectrum analyzer we used, and the high spatial resolution demanded. Spectral resolution was 0.07 nm and the spectrum was observed from 1030 to 1070 nm.

## 2.3 Tuneable laser for modal decomposition

As the previous method is time consuming and optical alignments are delicate due to the injection into the fibre, we propose to alleviate these drawbacks by simply scanning the wavelength of the source and observing the near-field image of the fibre end using a camera, as presented in Fig. 2(b). Magnification can be done by simply using a microscope objective. The magnified near-field image of the fibre output is recorded every time the wavelength of the source is changed. Data analysis is then performed similarly to the standard $S^2$ imaging technique. Using our experimental setup, there is no need for careful alignments, and only one dimension (the wavelength) has to be scanned, thus offering a much faster characterization. Also, modal analysis of small-mode-area fibres such as highly nonlinear fibres is easily performed using large magnification.

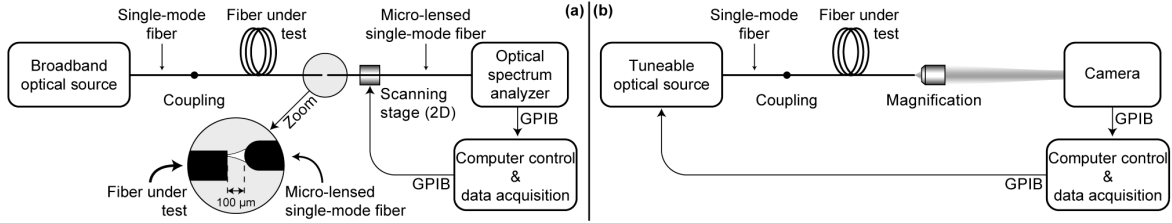

Fig. 2. Alternative experimental setups for modal decomposition of the fibre under test.

## 3. EXPERIMENTAL DEMONSTRATION

### 3.1 Microlensed fibre imaging for the $S^2$ technique

The experimental demonstration of the first technique has been done for an 80-cm-long photonic-bandgap fibre, referred to as Fibre A, specially designed at 1064 nm to minimize non-linear effects[7]. The single-mode output port of a commercial broadband optical source is spliced to Fibre A. The mode diameter is about 6 µm for the single-mode fibre and about 20 µm (averaged) for the fundamental mode of Fibre A, thus offering higher-order mode coupling due to mode-size mismatch at coupling. As shown in Fig. 3, a good agreement is obtained between experimental and theoretical modes. This technique could be used to optimize fibre coupling, e.g., using Gradissimo[8] fibres, so that only the fundamental mode of Fiber A is excited.

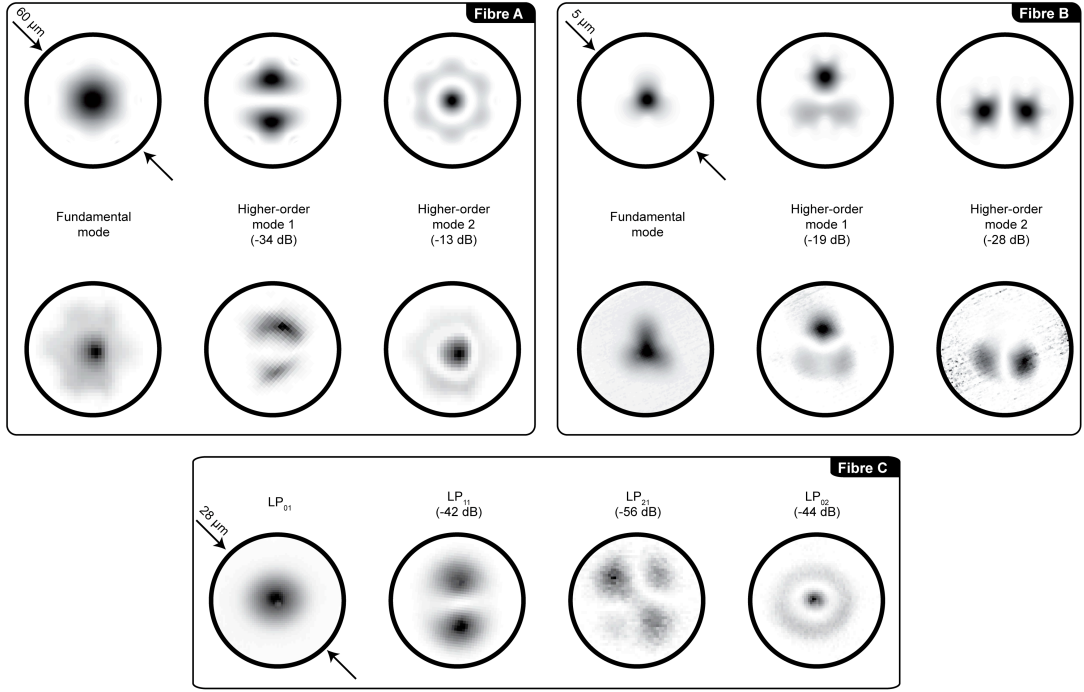

Fig. 3. Numerical (top) and experimental (bottom) intensity profiles of LP modes for Fibres A & B, experimental profiles for Fibre C.

### 3.2 Tuneable laser for modal decomposition

For the experimental demonstration of the second technique, we used a commercial C-band external-cavity laser. Light propagates in a standard single-mode fibre whose mode-field diameter is about 10.4 μm. This fibre is butt-coupled or spliced to the fibre under test. An 80 times microscope objective is used for magnification of the near-field images at the fibre output. Images are captured using a camera, which returns an 8-bits 200x200-points image. The experimental data acquisition was done in about ten minutes, thus offering great environment stability. Data were then processed in a few minutes. Two optical fibres, referred to as Fiber B and Fiber C, have been tested using this technique.

Fiber C is a 1-m-long standard stepped-index multi-mode fibre. The core diameter is 12 μm, and the core/cladding index difference is about $9.8 \cdot 10^{-3}$. Therefore, this fibre should support four LP modes: $LP_{01}$, $LP_{11}$, $LP_{02}$ and $LP_{21}$. These modes have been observed experimentally as shown in Fig. 3 by coiling the fibre with a 20 cm radius. We note that the $LP_{21}$ mode was not observed if the fibre was straight.

Fiber B is a 1.5-meter-long commercially-available solid-core photonic-crystal fibre whose fundamental mode has a 2.8-μm-average-diameter triangular shape. The laser wavelength is swept from 1544 nm every 0.01 nm over 5.12 nm. Experimental results are shown in Fig. 3. The fundamental mode is observed, as well as two higher-order modes. Optical power mainly resides in the fundamental mode. The most powerful higher-order mode is about 100 times less powerful than the fundamental mode. These optical fibres are often considered as single-mode fibres due to high propagation losses of the higher-order modes. However, multi-mode propagation can be deleterious for short fibres in some specific applications. Theoretical modes of the fibre were also calculated, and a good agreement was found with experimental results only if a transverse index inhomogeneity was included in the fibre structure. For a perfectly homogeneous fibre, higher-order modes 1 and 2 have the same intensity profiles, although they do have different effective indexes, but could not be distinguished experimentally.

Finally, we obtained similar results by replacing the expensive tuneable source by the combination of a cheap superfluorescent source and a tuneable optical filter with a 0.27-nm spectral width.

## 4. CONCLUSION

New methods for modal analysis of optical fibres were proposed and demonstrated. The first method uses a microlensed fibre for the $S^2$ imaging technique, thus avoiding the use of a magnification setup. This method has been used to characterize a photonic-bandgap large-mode-area fibre. The main drawback of this method along with existing ones is that it is time-consuming and requires careful optical alignments for the 2D scan. The second method is proposed and demonstrated to alleviate most of the drawbacks of existing techniques. Indeed, we proposed a simple and fast characterization process, which is suitable to any optical fibre and not only limited to large-mode-area fibres. This setup has been used to characterize a small-core microstructured optical fibre, and a standard index-stepped multi-mode fibre. We believe that this technique is a great tool for characterization of optical fibres, to control the condition of optical injection in a multi-mode fibre, to control the modal content by coiling the fibre, or to measure the propagation loss of the existing modes by combining this technique to cutback measurements.

## AKNOWLEDGEMENT

Authors are thankful to the *Région Bretagne* and the *Agence Nationale de la Recherche* for funding, respectively, the *Hippocamp* and *Futur* research programs.